# A Supervisory Frequency Support Control Scheme for Distributed PV

Qinmiao Li, *Member, IEEE*, Mesut Baran, *Fellow, IEEE*

***Abstract*—Increasing penetration of Photovoltaic (PV) generation brings an opportunity, and sometimes necessity, for this new resource to provide ancillary services such as frequency support. Recent efforts toward this goal focused mainly on the large-scale PV plants with identical subsystems but cannot handle distributed PV systems with diversities. Therefore, directly applying them to distributed PV requires repetitive control design for each unit, and thus implies huge design efforts. In this paper, we propose a novel frequency support control scheme focusing on the distributed PV to overcome the above limitation. Considering the diversities, we first derive a reduced-order aggregate model to represent the overall dynamic behaviors of a group of distributed PV. Using this model, we can then design one single frequency support controller for a large number of distributed PV without excessive design computational efforts. This controller aims at controlling the PV total output to provide frequency support to the grid. We also supplement it with the proposed inversion method to obtain individual PV's control signals from the aggregate one. The proposed reduced-order aggregate model is validated against a group of distributed PV systems represented by the detailed nonlinear models. We also demonstrate the effectiveness of proposed control scheme, including both controller and the inversion method, through time-domain simulations using a standard test system.**

## I. Introduction

PHOTOVOLTAIC (PV) generation has been fast growing in recent years [1], [2]. A considerable portion of the increasing PV installations are the small-size distributed PV in distribution systems which are projected to continue growing [3]. As of 2020, 37.3% of the total installed PV capacity in the United States is distributed PV [4]. PV systems are usually inverter-based systems without rotating inertia and governor systems, meaning that they do not have any frequency support capability. As a result, high PV penetration poses negative impacts on the bulk system frequency response such as low frequency nadir and large rate of change of frequency (RoCoF) [5]-[7]. Recently, several grid codes or orders from system operators and regulatory authorities started to require commercial PV plants to provide frequency support during system disturbances [8]-[11]. Meanwhile, leveraging distributed PV for grid services has become imperative, especially for small-scale power systems with high penetration of distributed PV. For example, the Hawaiian Electric is seeking frequency support capabilities from aggregators of distributed rooftop PV [12].

Methods for enabling frequency support from PV systems can be found in the literature. One approach is referred to as de-loading control [13], in which the PV's operating point is first set below its maximum power point (MPP). Then the PV array's voltage is controlled to adjust the output power in response of frequency changes [14]. Following this idea, authors in [15], [16] proposed frequency-droop controllers by using Newton quadratic interpolation and a lookup-table (LUT) based approach, respectively. In addition, virtual inertia control and frequency damping control are proposed in [17]. Another method for providing frequency support from PV systems is to utilize the energy stored in DC-link capacitors, and this can be achieved by adjusting the DC-link voltages to emulate inertial power response, as presented in [18]. Reference [19] and [20] also investigate the potential of combining the above two methods.

The aforementioned methods usually consider one single PV system or the large-scale PV plant that consists of identical subsystems, and thus do not endogenously support distributed PV with diversities. Adopting these methods and repeating the design process for each individual PV unit is theoretically feasible but will incur great control design effort as the number of PV units in a group is commonly very large. Moreover, having distributed PV managed by aggregators (at community or distribution level) is easier in practice and more effective for providing frequency support considering the scale effect. The recent Federal Energy Regulatory Commission (FERC) Order 2222 is also promoting the distributed PV to participate in ancillary services in aggregation [21]. Therefore, to leverage distributed PV for frequency support, a preferable solution is to aggregate them together and design one single controller whose design computational complexity will not be significantly high considering the large number of PV units. Following this idea, a fuzzy controller is proposed in [22] to regulate system frequency using distributed PV. However, the PV system model is very simplified by neglecting the local control dynamics, and only the insolation diversity among PV systems is considered.

In the above approach, apart from dealing with the variations in capacities, control parameters, and working conditions of distributed PV, another challenge is the modeling complexity for aggregating the distributed PV. The reason is that, even though the repetitive control design can be avoided, any aggregate model whose order is positively correlated to the number of PV units can still lead to great computational complexity. Hence, we first propose a new reduced-order aggregate model which can represent the overall dynamic behaviors of a group of distributed PV. The proposed model considers the diversities among individual PV, as well as retaining the order as that of one single PV system model. Then, using this model, we improve and adapt the frequency-tracking controller from our previous work [20] to the case of distributed PV for supervisory frequency support. The main motivation for choosing this method is its novelty of assuring the overall system frequency response (described by equivalent system inertia and droop) to be as desired. In the proposed overall control scheme (shown in Fig. 1), we also propose an inversion method as a supplement to the controller, so that the aggregate control signals based on the aggregate model can be inverted to each individual PV.

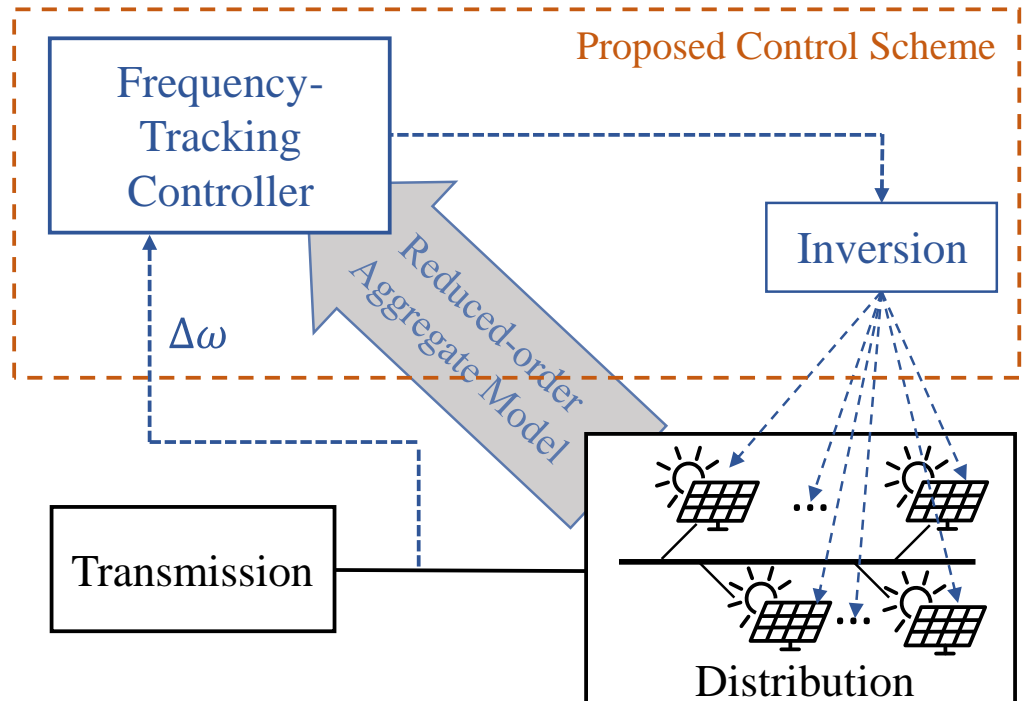

Fig. 1. Conceptual diagram of proposed control scheme.

In the literature, however, there is limited work on the aggregate PV system models, nor for reduced-order ones. An aggregate model for distributed PV with synchronous power controllers is proposed in [23]. However, the electrical part of PV system is modeled as just a current source with a first-order filter. Considering full components, study in [24] derives an aggregate model but only for multiple identical PV systems. For the case with diverse parameters, order reduction techniques are only suggested but no further details or examples are provided. In this paper, we derive a reduced-order aggregate model which addresses the above shortcomings. The obtained aggregate model is very similar to the small-signal model (SSM) for one PV system and has the same order. This can greatly reduce the computational effort. In addition, the model variables still have physical meanings that can facilitate control design and its implementation to a real system.

Hence, the contributions of this paper are: i) a reduced-order aggregate model that can represent the dynamics of a group of distributed PV with diversities; ii) an aggregate-model-based frequency support control scheme for distributed PVs without excessive design computational complexity. Section II of the paper introduces the new aggregate model. The proposed control scheme and test results are presented in Section III and IV. Conclusions are in Section V.

## II. The Reduced-order Aggregate Model for Distributed PV

In this section, we first introduce the SSM for one single-phase PV system. Then we derive its aggregate version for a group of distributed PV with different power ratings, control parameters, and solar irradiation.

### A. The Single-phase PV System

For one of the single-phase distributed PV, we consider it as a typical two-stage PV system, which is illustrated in Fig. 2. For such a PV system, existing models are liable to different limitations when they are considered for designing supervisory frequency support controller. For example, some models make simplifications by ignoring PV array characteristics and DC-link dynamics [25]-[27]. Another limitation is the lack of supervisory control input in the models, without which the output power cannot be adjusted in response of frequency[28]-[30]. The SSM proposed in [18] avoids above shortcomings, but it is derived for three-phase PV systems while distributed PV, especially rooftop PV, are mostly single-phase.

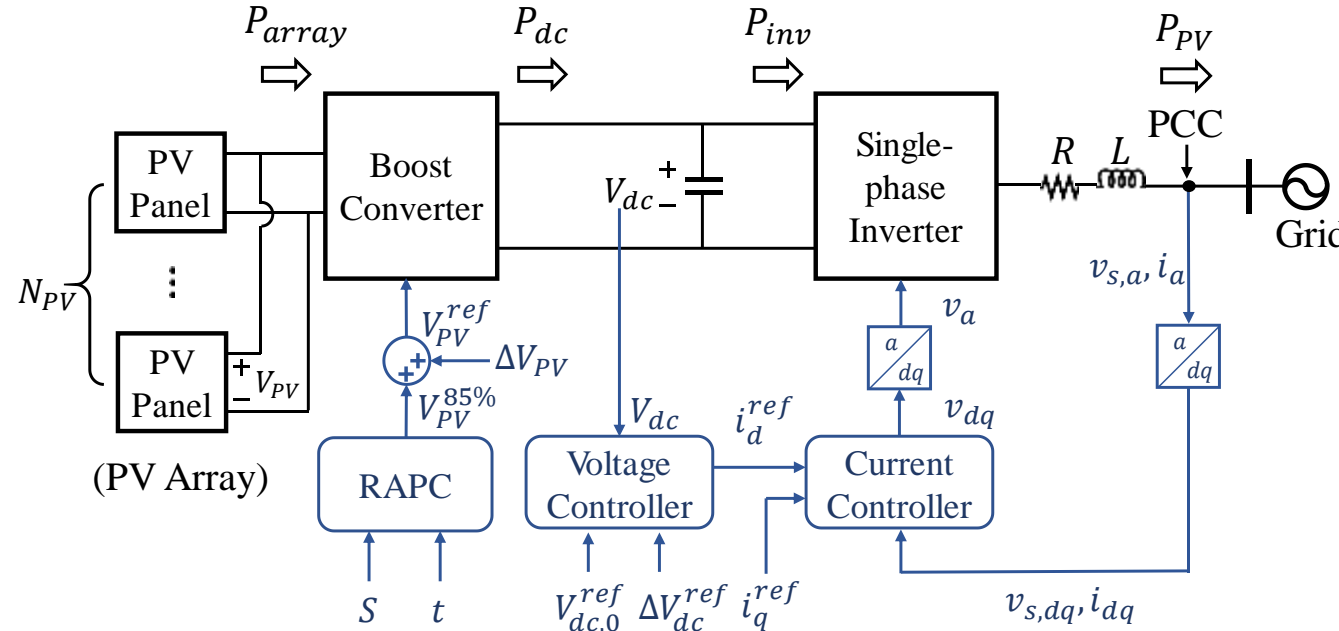

Fig. 2. Diagram for one single-phase two-stage PV system.

For the electrical part in Fig. 2, we assume the PV array consists of $N_{PV}$ identical PV panels connected in parallel. Its terminal voltage $V_{PV}$ is raised by a boost converter. The DC-link serving as the energy buffer is in between of the converter and a single-phase inverter. An $R-L$ filter is used to interface with the grid. For the local control of this PV system, the rapid active power control (RAPC) method [16] is used at boost converter to determine the operating point of PV array, and thus adjust the PV power output. A revised dual-loop current mode controller with feedforward compensation [28] is adopted for inverter control. $\Delta V_{PV}$ and $\Delta V_{dc}^{ref}$ are two control input that we added to the above local controllers for designing supervisory frequency support control functions.

Compared to the three-phase PV system in [18], the single-phase PV system shares the same topology and local control scheme. The only difference that we need to address is the power balance equation:

$$P_{PV} = \frac{1}{2}\left(i_d v_{sd} + i_q v_{sq}\right) = \frac{1}{2} i_d v_{sd} \tag{1}$$

in which $i_d$, $i_q$ and $v_{sd}$, $v_{sq}$ are the $dq$ components of current and voltage at point of common coupling (PCC), respectively. They can be obtained by the single-phase $dq$ transformation based on the system voltage ($v_{s,a}$) [31]. The $\omega$ and $\theta$ used in the transformation are provided by an ideal phase-locked loop (PLL). The PLL also regulates $v_{sq}$ to zero, so that the simplification in (1) can be made.

Then, we can proceed the derivation similarly as the three-phase case in [18] to model the rest of the system, and obtain the SSM for the single-phase two-stage PV system in Fig. 2:

$$\frac{d\Delta V_{dc}}{dt} = \frac{1}{CV_{dc,0}^{ref}}\big(\underbrace{N_{PV}\, g(\Delta V_{PV}, S, t)}_{\Delta P_{array}} - \underbrace{\frac{1}{2}\Delta i_d V_{sd}}_{\Delta P_{PV}}\big) \tag{2}$$

$$\begin{aligned}\frac{d\Delta P_{PV}}{dt} = -\frac{1}{\tau}\Delta P_{PV} - \frac{k_P V_{sd}}{2\tau}\Delta V_{dc} + \frac{k_P V_{sd}}{2\tau}\Delta V_{dc}^{ref} \\ + \frac{V_{sd}}{2\tau}\Delta x\end{aligned} \tag{3}$$

$$\frac{d\Delta x}{dt} = -k_I \Delta V_{dc} + k_I \Delta V_{dc}^{ref} \tag{4}$$

where the function $g(\Delta V_{PV}, S, t)$ maps the change of PV array voltage to the change of PV panel's output power under given ambient conditions. This function is best represented by an LUT, as it is highly nonlinear. $V_{sd}$ is the nominal grid voltage which is assumed to be well-regulated by the grid. $k_P$ and $k_I$ are the parameters in voltage loop controller, and $\tau$ is the time constant in current loop controller [28]. $\Delta x$ represents the small-signal increment of a controller state which we introduced to avoid second-order derivative in the model.

### B. *The Reduced-order Aggregate Model*

Let $\mathcal{P} = \{1,2,3,\dots,N\}$ denote the set of distributed PV in an area of distribution system. For one PV system $i \in \mathcal{P}$, we have its SSM in the form of:

$$\frac{d\Delta V_{dc,i}}{dt} = \frac{\left(\frac{P_{r,i}}{P_{pa,i}} g_i\left(\Delta V_{PV,i}, S_i, t_i\right) - \Delta P_{PV,i}\right)}{C_i V_{dc,0,i}^{ref}} \tag{5}$$

$$\begin{aligned}\frac{d\Delta P_{PV,i}}{dt} &= -\frac{1}{\tau_i}\Delta P_{PV,i} - \frac{k_{P,i}V_{sd,i}}{2\tau_i}\Delta V_{dc,i} \\ &\quad + \frac{k_{P,i}V_{sd,i}}{2\tau_i}\Delta V_{dc,i}^{ref} + \frac{V_{sd,i}}{2\tau_i}\Delta x_i\end{aligned} \tag{6}$$

$$\frac{d\Delta x_i}{dt} = -k_{I,i}\Delta V_{dc,i} + k_{I,i}\Delta V_{dc,i}^{ref} \tag{7}$$

where $P_{r,i}$ and $P_{pa,i}$ are the rated power of the PV system and one PV panel, respectively, and

*a.1) PV Panel:* PV panels used in small PV systems usually have similar characteristics. Hence, we can assume that the PV panels are the same for all PV systems in $\mathcal{P}$, and thus, $\forall i \in \mathcal{P}$:

$$P_{pa,i} = P_{pa} \tag{8}$$

$$g_i\left(\Delta V_{PV,i}, S_i, t_i\right) = g\left(\Delta V_{PV,i}, S_i, t_i\right) \tag{9}$$

Also, since the PV systems are within the same area of distribution system, the ambient temperature for PV array does not vary a lot, i.e. $\forall i \in \mathcal{P}$: $t_i = t$. For solar irradiance, however, we keep the variance among each PV system because the cloud distribution can be notably different within the area.

*a.2) Control Parameters:* For the inverter controller which regulates the DC-link voltage, the voltage reference mainly depends on the inverter output voltage level. Since all the PV systems are in the same distribution system with the same voltage level, we can assume identical DC-link voltage references, and hence $\forall i \in \mathcal{P}$: $V_{dc,0,i}^{ref} = V_{dc,0}^{ref}$.

The other control parameter is the current control loop time constant $\tau_i$. This constant has to be very small such that the current loop is sufficiently fast to coordinate with outer voltage loop. Therefore, we can neglect the variations on $\tau_i$ and let $\tau_i = \tau$, because the outer voltage loop control parameters are more dominant in shaping the controller dynamics.

*a.3) DC-link Capacitor:* The main principle of choosing the DC-link capacitor value is to help stabilize the voltage such that its ripples are within a desired range [27], [32]. Following the design guide in [27], we have $\forall i \in \mathcal{P}$:

$$C_i = P_{r,i} \times \underbrace{\frac{D^{max} \times T_s^{max}}{\alpha\% \times \left(V_{dc,0}^{ref}\right)^2}}_{constant} = P_{r,i} \times C_d \tag{10}$$

where $D^{max}$ and $T_s^{max}$ are the maximum values of the boost converter's duty cycle and switching period, respectively, which we can assume to be the same for each PV system. The maximum voltage ripple level, $\alpha\%$, is also chosen to be the same. Moreover, if we assume the DC-link voltages to be regulated very close to the nominal value $V_{dc,0}^{ref}$, the capacitance $C_i$ for the $i$-th PV system becomes a product of the rated power, $P_{r,i}$, and a constant unit capacitance, $C_d$, as indicated in (10).

TABLE I
VARIABLE DISTRIBUTIONS AND CORRESPONDING PARAMETERS

| **MCS for LUT** | | |
|---|---|---|
| $P_{r,i}$ | $P_{r,i} \sim \mathcal{N}(\mu_P, \sigma_P^2)$ | $\mu_P \sim \mathcal{U}(150000, 250000)$, $\sigma_P \sim \mathcal{U}(10, 30)$ |
| $\Delta V_{PV,i}$ | $\Delta V_{PV,i} \sim \mathcal{N}(\mu_{V_{PV}}, \sigma_{V_{PV}}^2)$ | $\mu_{V_{PV}} \sim \mathcal{U}(-30, -7.5)$, $\sigma_{V_{PV}} \sim \mathcal{U}(2, 7)$ |
| $S_i$ | $S_i \sim \mathcal{N}(\mu_S, \sigma_S^2)$ | $\mu_S \sim \mathcal{U}(40, 70)$, $\sigma_S \sim \mathcal{U}(1, 10)$ |
| $N$ | $N \sim \mathcal{U}(N_{MIN}, N_{MAX})$ | $N_{MIN} = 10$, $N_{MAX} = 100$ |
| **MCS for $c_P$ and $c_I$** | | |
| $P_{r,i}$ | $P_{r,i} \sim \mathcal{N}(\mu_P, \sigma_P^2)$ | $\mu_P \sim \mathcal{U}(150000, 250000)$, $\sigma_P \sim \mathcal{U}(10, 30)$ |
| $V_{sd,i}$ | $V_{sd,i} \sim \mathcal{N}(\mu_{V_{sd}}, \sigma_{V_{sd}}^2)$ | $\mu_{V_{PV}} = -335$, $\sigma_{V_{PV}} = 5.6$ |
| $k_{P,i}$ | $k_{P,i} \sim \mathcal{U}(k_{P,MIN}, k_{P,MAX})$ | $k_{P,MIN} = 10$, $k_{P,MAX} = 100$ |
| $k_{I,i}$ | $k_{I,i} \sim \mathcal{U}(k_{I,MIN}, k_{I,MAX})$ | $k_{I,MIN} = 10$, $k_{I,MAX} = 100$ |
| $\lvert\mathcal{P}\rvert(N)$ | $N \sim \mathcal{U}(N_{MIN}, N_{MAX})$ | $N_{MIN} = 10$, $N_{MAX} = 100$ |

Next, to obtain the aggregate model, we first define some aggregate variables as:

$$P_r^a = \sum_{i\in\mathcal{P}} P_{r,i}, \quad \Delta P_{PV}^a = \sum_{i\in\mathcal{P}} \Delta P_{PV,i}, \quad \Delta x^a = \sum_{i\in\mathcal{P}} V_{sd,i}\Delta x_i$$

$$\Delta V_{dc}^a = \frac{\sum_{i\in\mathcal{P}} P_{r,i}\Delta V_{dc,i}}{\sum_{i\in\mathcal{P}} P_{r,i}}, \Delta V_{dc}^{ref,a} = \frac{\sum_{i\in\mathcal{P}} P_{r,i}\Delta V_{dc,i}^{ref}}{\sum_{i\in\mathcal{P}} P_{r,i}}$$

$$\Delta V_{PV}^a = \frac{\sum_{i\in\mathcal{P}} P_{r,i}\Delta V_{PV,i}}{\sum_{i\in\mathcal{P}} P_{r,i}}, \; S^a = \frac{\sum_{i\in\mathcal{P}} P_{r,i}S_i}{\sum_{i\in\mathcal{P}} P_{r,i}}$$

Considering the above assumptions, these aggregate variables allow us to obtain the following aggregate model by summing up the SSMs in (5)-(7) for all PV systems in $\mathcal{P}$:

$$\begin{aligned}\frac{d\Delta V_{dc}^a}{dt} &= \frac{1}{C_d V_{dc,0}^{ref} P_{pa}} \underbrace{\frac{\sum_{i\in\mathcal{P}} P_{r,i} g\left(\Delta V_{PV,i}, S_i, t\right)}{P_r^a}}_{S.1} \\ &\quad - \frac{1}{C_d V_{dc,0}^{ref} P_r^a}\Delta P_{PV}^a\end{aligned} \tag{11}$$

$$\begin{aligned}\frac{d\Delta P_{PV}^a}{dt} &= -\frac{1}{\tau}\Delta P_{PV}^a - \frac{1}{2\tau}\underbrace{\sum_{i\in\mathcal{P}} V_{sd,i}k_{P,i}\Delta V_{dc,i}}_{S.2} \\ &\quad + \frac{1}{2\tau}\underbrace{\sum_{i\in\mathcal{P}} V_{sd,i}k_{P,i}\Delta V_{dc,i}^{ref}}_{S.3} + \frac{1}{2\tau}\Delta x^a\end{aligned} \tag{12}$$

$$\frac{d\Delta x^a}{dt} = -\underbrace{\sum_{i\in\mathcal{P}} V_{sd,i}k_{I,i}\Delta V_{dc,i}}_{S.4} + \underbrace{\sum_{i\in\mathcal{P}} V_{sd,i}k_{I,i}\Delta V_{dc,i}^{ref}}_{S.5} \tag{13}$$

In (11)-(13), the summation terms (S.1-S.5) indicate that the model order has not been fully reduced. Therefore, we want to represent these summation terms with the aggregate variables to further reduce the order. In the following, we discuss how this is achieved for each term.

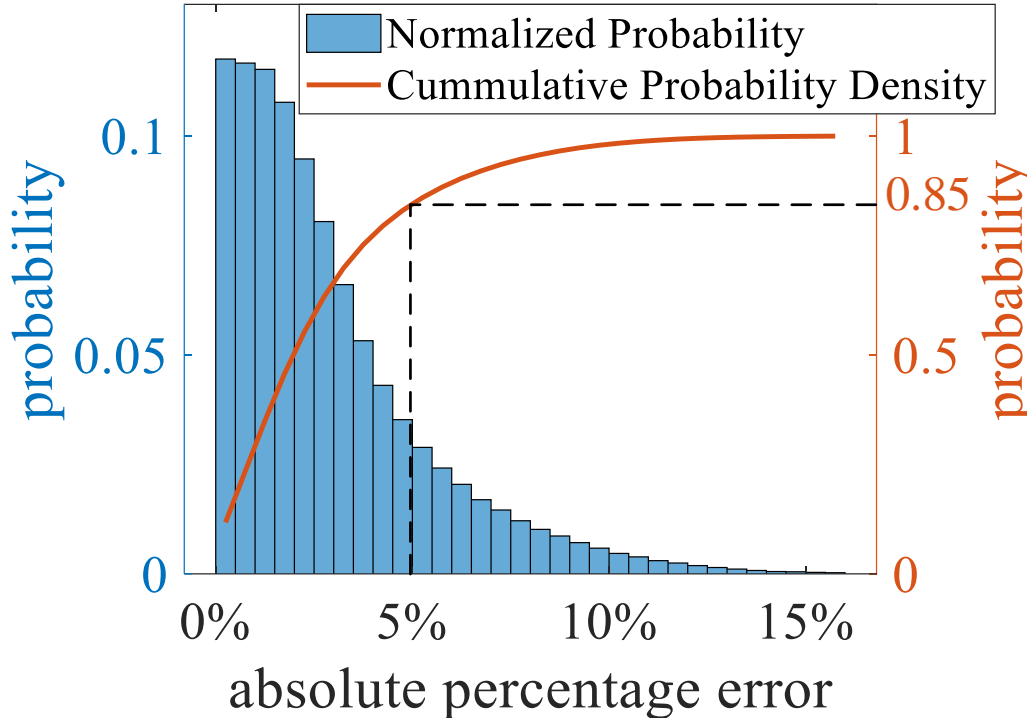

Fig. 3. Probability distribution and cumulative density of the LUT approximation error from MCS.

*b.1) S.1:* For this summation term, we approximate it as:

$$\frac{\sum_{i\in\mathcal{P}} P_{r,i}\, g\left(\Delta V_{PV,i}, S_i, t\right)}{P_r^a} \approx g\left(\Delta V_{PV}^a, S^a, t\right) \tag{14}$$

The idea of this approximation is to move the weighted averaging on LHS into $g$ on the variables $\Delta V_{PV,i}$ and $S_i$. Since the function $g$ represents an LUT, it is very difficult to analytically evaluate the approximation error. Therefore, we take the numerical approach of Monte Carlo Simulation (MCS) to assess the error level. For the MCS, we first assume that all the variables ($P_{r,i}$ , $\Delta V_{PV,i}$ and $S_i$) follow normal distributions with certain means and variances. Moreover, we consider the means and variances to be random samples from corresponding uniform distributions. Details about the distributions are summarized in TABLE I.

In the MCS, we set the number of trials to be 500,000. For each trial, we first determine $N$ (number of distributed PV) as a random sample from $[10, 100]$, and then sample $N$ sets of the three variables to compute the approximation error of (14). Fig. 3 shows the probability distribution of calculated absolute percentage error, as well as the cumulative probability density curve. From the plot, the error is very small for most of the cases, and we have a probability of 85% to get an error below 5%. Therefore, (14) represents a good approximation with satisfactory accuracy.

*b.2) S.2-S.5:* Summation terms S.2-S.5 are discussed together because they share the same structure. The same treatment can be applied to all of them. Taking S.2 as an example, we can rearrange it as:

$$\begin{aligned}\sum_{i\in\mathcal{P}} V_{sd,i} k_{P,i} \Delta V_{dc,i} &= \sum_{i\in\mathcal{P}} \frac{V_{sd,i} k_{P,i}}{P_{r,i}} P_r^a \Delta V_{dc}^a \\ &\quad - \sum_{i\in\mathcal{P}} \left( \sum_{j\in\mathcal{P}, j\neq i} \frac{V_{sd,j} k_{P,j}}{P_{r,j}} \right) P_{r,i} \Delta V_{dc,i}\end{aligned} \tag{15}$$

Note that the summation terms can be further simplified if the following holds:

$$\frac{V_{sd,i} k_{P,i}}{P_{r,i}} = c_P, \forall i \in \mathcal{P} \tag{16}$$

where $c_P$ is a constant parameter. Equation (16) may not hold in practice, but a good value for $c_P$ can be obtained by minimizing the Euclidean norm of error between $\frac{V_{sd,i} k_{P,i}}{P_{r,i}}$ and $c_P$. To achieve this, we may construct an optimization problem to determine $c_P$ as follows:

$$\underset{c_P}{Minimize} \sum_{i\in\mathcal{P}} \left( c_P - \frac{V_{sd,i} k_{P,i}}{P_{r,i}} \right)^2 \tag{17}$$

Solving the optimization problem, we can obtain:

$$c_P = \frac{\sum_{i\in\mathcal{P}} \left( \frac{V_{sd,i} k_{P,i}}{P_{r,i}} \right)}{|\mathcal{P}|} \tag{18}$$

where $|\mathcal{P}|$ is the cardinality of set $\mathcal{P}$. Substitute (16) into (15) to obtain:

$$\sum_{i\in\mathcal{P}} V_{sd,i} k_{P,i} \Delta V_{dc,i} \approx c_P P_r^a \Delta V_{dc}^a \tag{19}$$

where S.2 is successfully expressed by the aggregate variables.

The same treatment can also be applied on S.3-S.5 to have:

$$\sum_{i\in\mathcal{P}} V_{sd,i} k_{P,i} \Delta V_{dc,i}^{ref} \approx c_P P_r^a \Delta V_{dc}^{ref,a} \tag{20}$$

$$\sum_{i\in\mathcal{P}} V_{sd,i} k_{I,i} \Delta V_{dc,i} \approx c_I P_r^a \Delta V_{dc}^a \tag{21}$$

$$\sum_{i\in\mathcal{P}} V_{sd,i} k_{I,i} \Delta V_{dc,i}^{ref} \approx c_I P_r^a \Delta V_{dc}^{ref,a} \tag{22}$$

where

$$c_I = \frac{\sum_{i\in\mathcal{P}} \left( \frac{V_{sd,i} k_{I,i}}{P_{r,i}} \right)}{|\mathcal{P}|} \tag{23}$$

Note that $V_{sd,i}$, which is the grid-side voltage at every node of a distribution network, should be well-regulated around $\pm 5\%$ from its nominal value $V_{sd}$. Moreover, the value of $c_P$ and $c_I$ are mainly affected by $P_{r,i}$ which is commonly in the order of $10^4$ or $10^5$ (e.g. 200 KW), while $V_{sd,i}$, $k_{P,i}$, and $k_{I,i}$ are in the order of only 10 or $10^2$. Therefore, neglecting the variances by using $V_{sd}$ in replace of $V_{sd,i}$ should not cause significant errors of $c_P$ and $c_I$, and thus, we have (18) and (23) as:

$$c_P \approx \frac{V_{sd} \sum_{i\in\mathcal{P}} \left( \frac{k_{P,i}}{P_{r,i}} \right)}{|\mathcal{P}|} = V_{sd} c_P' \tag{24}$$

$$c_I \approx \frac{V_{sd} \sum_{i\in\mathcal{P}} \left( \frac{k_{I,i}}{P_{r,i}} \right)}{|\mathcal{P}|} = V_{sd} c_I' \tag{25}$$

To validate the above approximations, we again conduct the MCS to evaluate. We assume that $V_{sd,i}$ follows the normal distribution with mean $V_{sd}$, and the standard deviation is chosen to be $1.67\% V_{sd}$, so that nearly all the $V_{sd,i}$ samples fall within $\pm 5\% V_{sd}$. $P_{r,i}$ follows the same normal distribution as in b.1). The rest parameters ($k_{P,i}$, $k_{I,i}$ and $|\mathcal{P}|$) follow corresponding uniform distributions. Details about the parameter distributions are also summarized in TABLE I.

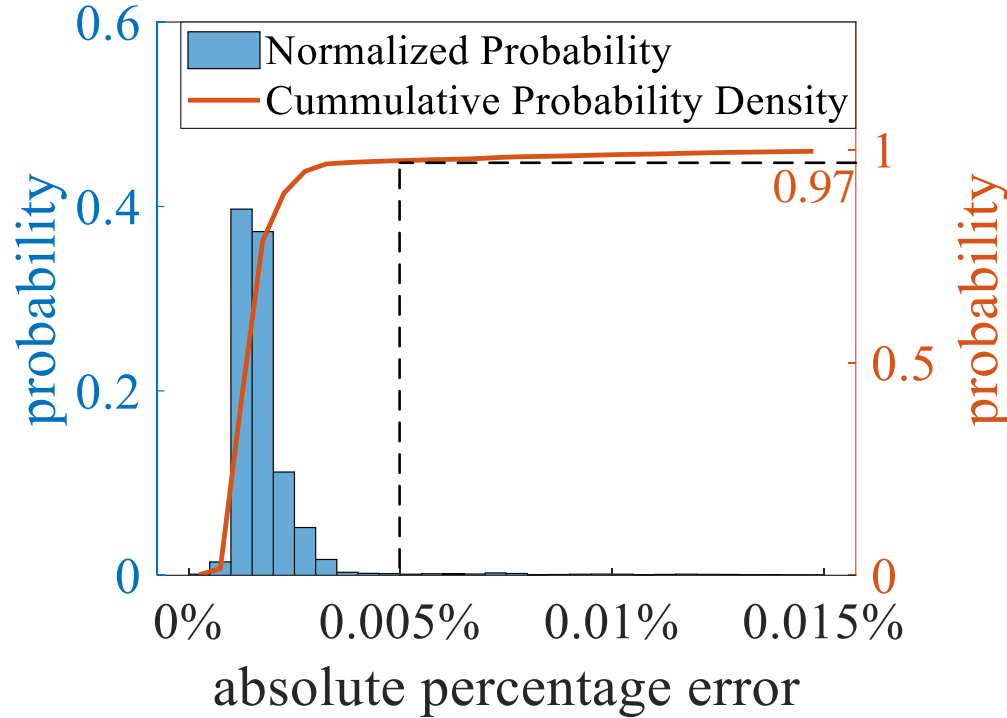

Fig. 4. Probability distribution and cumulative density of the $c_P$ and $c_I$ approximation error from MCS.

For the MCS, we run 500,000 trials respectively for $c_P$ and $c_I$, and calculate the approximation error of (24) and (25). Fig. 4 summarizes the probability distribution and cumulative density of the calculated absolute percentage error. The results indicate very good performance with a probability of 97% to have an absolute percentage error below 0.005%.

Now we have the reduced-order aggregate model as:

$$\frac{d\Delta V_{dc}^{a}}{dt}=\frac{1}{C_d V_{dc,0}^{ref}}\left(\frac{g(\Delta V_{PV}^{a},S^{a},t)}{P_{pa}}-\frac{\Delta P_{PV}^{a}}{P_r^{a}}\right) \quad (26)$$

$$\frac{d\Delta P_{PV}^{a}}{dt}=-\frac{1}{\tau}\Delta P_{PV}^{a}-\frac{V_{sd}}{2\tau}c'_P P_r^{a}\Delta V_{dc}^{a}+\frac{V_{sd}}{2\tau}c'_P P_r^{a}\Delta V_{dc}^{ref,a}+\frac{1}{2\tau}\Delta x^{a} \quad (27)$$

$$\frac{d\Delta x^{a}}{dt}=-V_{sd}c'_I P_r^{a}\Delta V_{dc}^{a}+V_{sd}c'_I P_r^{a}\Delta V_{dc}^{ref,a} \quad (28)$$

Note that this aggregate model is similar to the SSM for one PV system (2)-(4) with the same order. Thus, this model allows us to view the collective dynamics of a group of distributed PV as that of one large-scale PV system, and then design supervisory control for it. Furthermore, the aggregate state and input variables still correspond to physical quantities, such as the total increment power generation of the distributed PV $\Delta P_{PV}^{a}$. This facilitates the later control design in Section III.

## III. Supervisory Control Scheme for Frequency

In this section, we introduce the proposed control scheme in which the frequency-tracking controller from [20] is improved based on the reduced-order aggregate model with an inversion function supplemented, as illustrated in Fig. 1. The frequency-tracking controller consists of an LUT and an unknown input observer (UIO) based tracking linear quadratic regulator (LQR), which are detailed in Fig. 5. The role of this LUT is to compensate the nonlinearity in the reduced-order aggregate model, and thus obtain a linear combined system model. Taking this linear model as the plant model, we can then design the tracking LQR to control the output of distributed PV such that the overall system frequency can effectively track that of a designed reference system with desired inertia and droop constants ($H^{ref}$ and $R^{ref}$ ). In other words, with this controller, the desired frequency response capability of the close-loop system (including distributed PV) can be achieved. At last, the inversion function inverts the aggregate control signals, which corresponds to the aggregate model, to individual ones for each distributed PV.

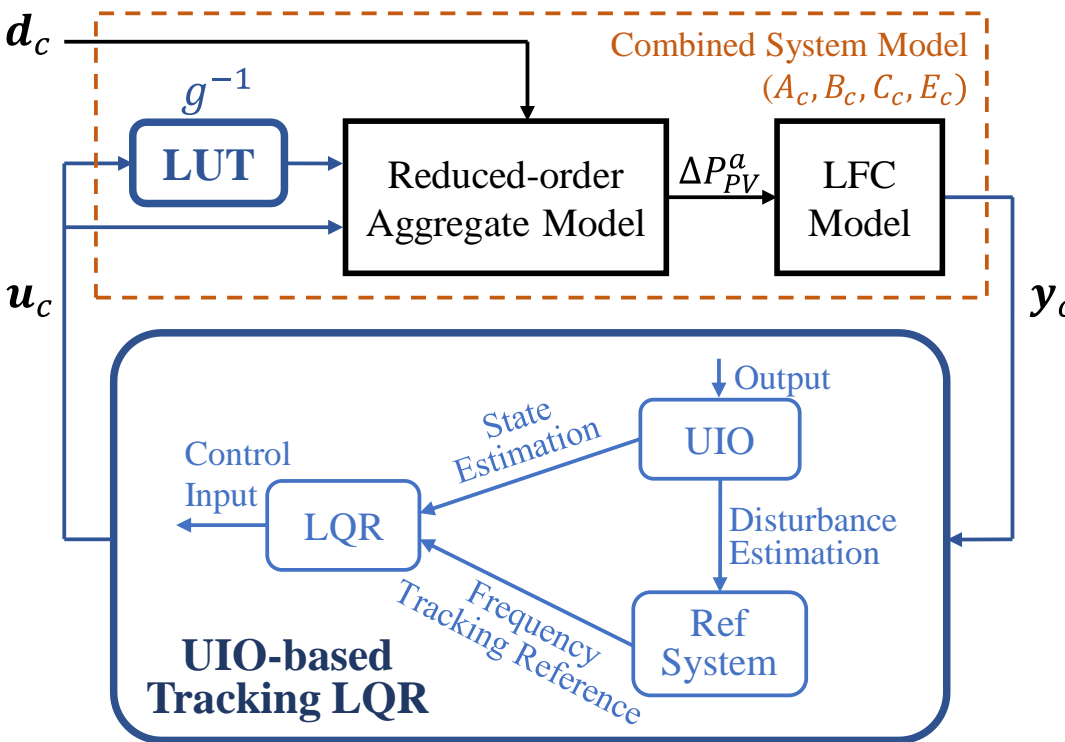

Fig. 5. Architecture and components of the frequency-tracking controller

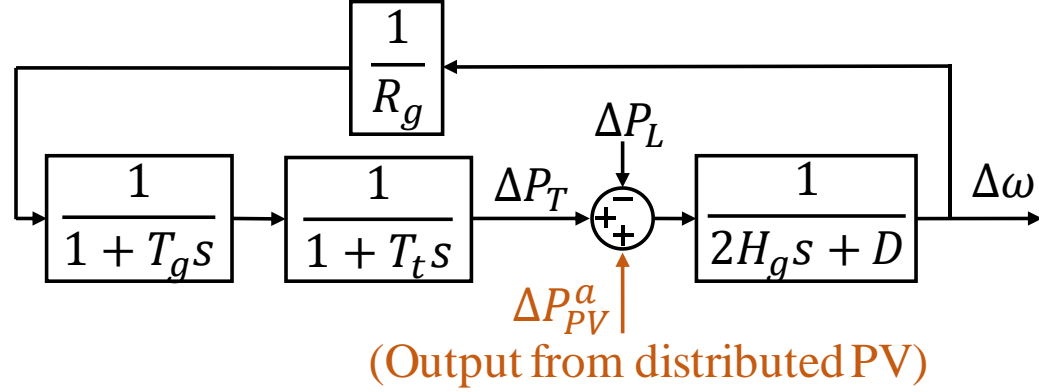

Fig. 6. LFC model including distributed PV

This control scheme is similar to the one in [20]. However, since the later one considers a large-scale PV plant with a group of identical PV subsystems, applying its method to distributed PV with diversities requires repetitive control designs based on each PV system's model. As a result, the complete design for distributed PV implies $\mathcal{O}(N)$ computational complexity, where $N$ as the number of PV units can be in the order up to thousands. Instead, benefited from the reduced-order aggregate model, our proposed control scheme only requires $\mathcal{O}(1)$ computational complexity, which is constant and decoupled from $N$. The other improvement in this work against [20] is the inversion method that fills the implementation gap between a central controller and distributed PV by inverting the aggregate control signals to each PV unit.

### A. LUT and The Combined System Model

As mentioned earlier, the LUT is introduced to eventually obtain a linear combined system model as the plant model for designing the tracking LQR. In this regard, we first use the classical load frequency control (LFC) model to describe the real power and frequency dynamics in the transmission system. Fig. 6 illustrates the diagram of this model, and how the distributed PV can be tied to this model. From Fig. 6, we can write the LFC model in state-space form as:

$$\begin{cases}\dot{\boldsymbol{x}}_g=\boldsymbol{A}_g\boldsymbol{x}_g+\boldsymbol{B}_g\boldsymbol{u}_g+\boldsymbol{E}_g\boldsymbol{d}_g\\ \boldsymbol{y}_g=\boldsymbol{C}_g\boldsymbol{x}_g\end{cases} \quad (29)$$

where we treat the increment power $\Delta P_{PV}^{a}$ from distributed PV as an input ($\boldsymbol{u}_g$) to the transmission system whereas the load change $\Delta P_L$ is taken as the disturbance ($\boldsymbol{d}_g$).

Then, for the distributed PV, its derived reduced-order aggregate model in (26)-(28) still contains a nonlinear term which is the LUT function $g(\Delta V_{PV}^{a},S^{a},t)$. However, the LQR design will be facilitated if the plant (combined system) model is linear. Therefore, we add another LUT block between the

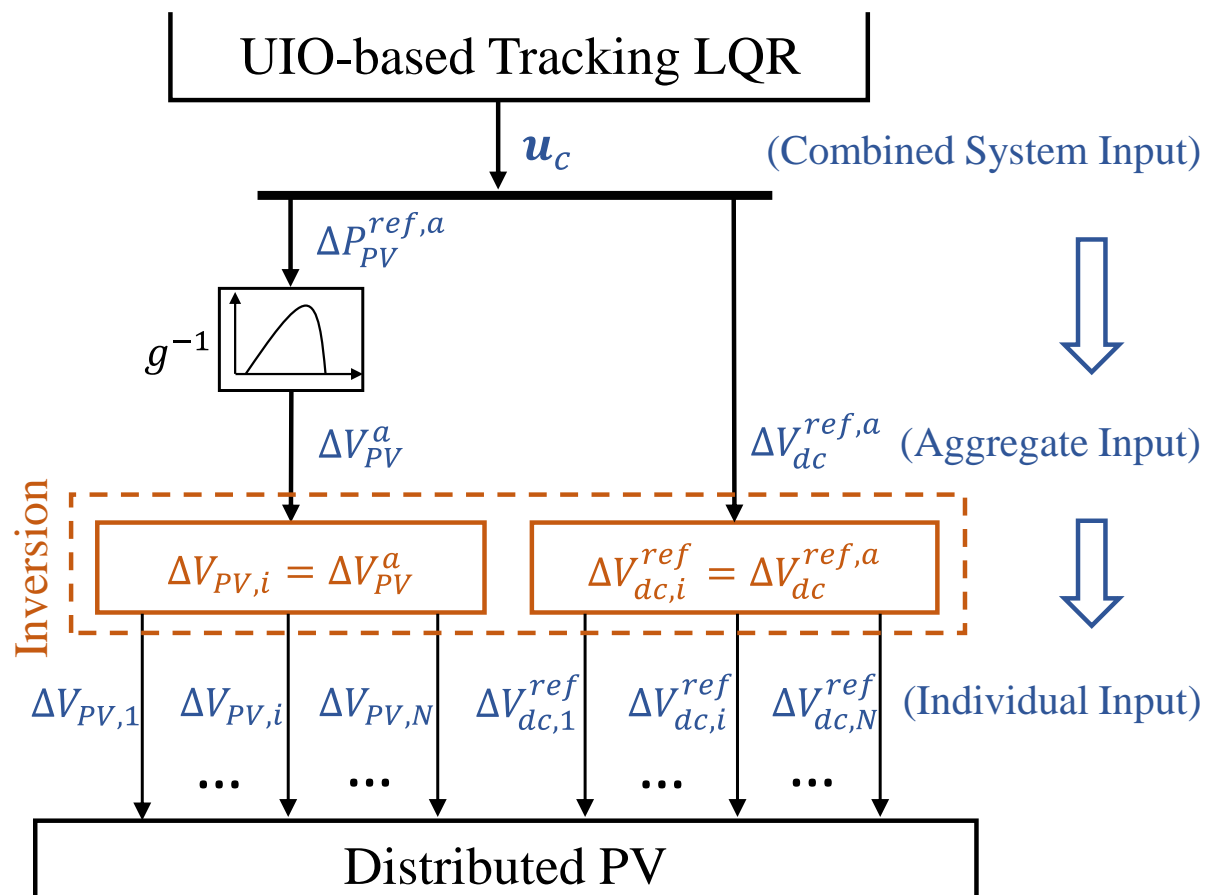

Fig. 7. Inversion of control signals

tracking LQR and distributed PV, as shown in Fig. 1. If this LUT is chosen as the inverse of $g$:

$$\Delta V_{PV}^{a} = g^{-1}(\Delta P_{PV}^{ref,a}, S^{a}, t) \tag{30}$$

The plant to be seen by the tracking LQR becomes a linear model by substituting (30) into (26). This linear model can be written in the state-space form as:

$$\begin{cases} \dot{\boldsymbol{x}}_{PV} = \boldsymbol{A}_{PV}\boldsymbol{x}_{PV} + \boldsymbol{B}_{PV}\boldsymbol{u}_{PV} \\ \boldsymbol{y}_{PV} = \boldsymbol{C}_{PV}\boldsymbol{x}_{PV} \end{cases} \tag{31}$$

where $\boldsymbol{x}_{PV} = [\Delta V_{dc}^{a}\ \ \Delta P_{PV}^{a}\ \ \Delta x^{a}]^{T}$, $\boldsymbol{u}_{PV} = \left[\Delta P_{PV}^{ref,a}\ \ \Delta V_{dc}^{ref,a}\right]^{T}$, and $\boldsymbol{y}_{PV} = [\Delta P_{PV}^{a}]$.

Now we can combine the two linear models for both transmission system (29) and distributed PV (with an LUT) (31) through the interfacing variable $\Delta P_{PV}^{a}$. The obtained model is referred as the combined system model, which is used as the plant model for the following design of tracking LQR. The combined system model can also be described in state-space as:

$$\underbrace{\begin{bmatrix} \dot{\boldsymbol{x}}_g \\ \dot{\boldsymbol{x}}_{PV} \end{bmatrix}}_{\dot{\boldsymbol{x}}_c} = \underbrace{\begin{bmatrix} \boldsymbol{A}_g & \boldsymbol{B}_g\boldsymbol{C}_{PV} \\ 0 & \boldsymbol{A}_{PV} \end{bmatrix}}_{\boldsymbol{A}_c} \underbrace{\begin{bmatrix} \boldsymbol{x}_g \\ \boldsymbol{x}_{PV} \end{bmatrix}}_{\boldsymbol{x}_c} + \underbrace{\begin{bmatrix} 0 \\ \boldsymbol{B}_{PV} \end{bmatrix}}_{\boldsymbol{B}_c} \underbrace{[\boldsymbol{u}_{PV}]}_{\boldsymbol{u}_c} + \underbrace{\begin{bmatrix} \boldsymbol{E}_g \\ 0 \end{bmatrix}}_{\boldsymbol{E}_c} \underbrace{[\boldsymbol{d}_g]}_{\boldsymbol{d}_c} \tag{32}$$

$$\boldsymbol{y}_c = \boldsymbol{C}_c\boldsymbol{x}_c \tag{33}$$

where the output $\boldsymbol{y}_c$ is frequency deviation $\Delta\omega$ and the input $\boldsymbol{u}_c$ is actually the input for distributed PV ($\boldsymbol{u}_{PV}$).

### B. The UIO-based Tracking LQR

Based on the linear combined system model, the UIO-based tracking LQR can be designed to let the system frequency effectively track that of a reference system under the same unknown load disturbance. To achieve this objective, three components are designed, namely, the reference system, the UIO, and the LQR, as illustrated by Fig. 5. The reference system, as suggested by name, serves as a "desired version" of the real transmission system. It is a virtual system in the form of LFC model whose inertia and droop are designed to be our desired values ($H^{ref}$ and $R^{ref}$). In this way, under the same disturbance, its frequency response can be used as a reference for the real system with distributed PV to track. However, the disturbance occurred to the real system cannot be known and provided to the reference system in advance. Therefore, to address this difficulty, a UIO is used to estimate the unknown disturbance, as well as the system states, in time [33], [34]. The estimated disturbance is sent to the reference system to generate the reference frequency response, while the state estimation is used by the LQR. As the last component, the LQR follows a standard design, where the frequency tracking error (calculated from real system frequency and the reference system frequency) is included in the objective function, and to be minimized. Further details about the design can be found in [20].

### C. Control Signal Inversion

Since the controller is designed based on the combined system, the control input should also correspond to the input ($\boldsymbol{u}_c$) of the combined system model. From (32), $\boldsymbol{u}_c$ is actually $\boldsymbol{u}_{PV}$ which comprises two inputs: $\Delta P_{PV}^{ref,a}$ and $\Delta V_{dc}^{ref,a}$, where $\Delta P_{PV}^{ref,a}$ can be converted to $\Delta V_{PV}^{a}$ by the LUT function $g^{-1}$. So eventually, the control signals $\boldsymbol{u}_c$ are mapped to aggregate inputs ($\boldsymbol{u}^{a} = \left[\Delta V_{PV}^{a}\ \ \Delta V_{dc}^{ref,a}\right]^{T}$) in the aggregate PV model (see Fig. 7). However, for practical implementation, we need the individual control input ( $\boldsymbol{u}_i = \left[\Delta V_{PV,i}\ \ \Delta V_{dc,i}^{ref}\right]^{T}$ ) for each distributed PV.

Note that, however, there is no unique way of inverting $\boldsymbol{u}^{a}$ back to $\boldsymbol{u}_i$. In principle, any inversion should work, as long as satisfying the one-way mapping from $\boldsymbol{u}_i$ to $\boldsymbol{u}^{a}$ as described in Section II.B. Since the individual control input ($\boldsymbol{u}_i$) determines how much PV system's power output will change, a good strategy is to have the PV systems with higher capability, in terms of rated power or irradiance, to contribute more power. Following this idea, we propose the inversion rule as:

$$\Delta V_{PV,i} = \Delta V_{PV}^{a}, \qquad \Delta V_{dc,i}^{ref} = \Delta V_{dc}^{ref,a} \tag{34}$$

which implies setting all the individual control inputs equal to the aggregate input. This inversion is also illustrated in Fig. 7. First, we can easily verify that the mapping from $\boldsymbol{u}_i$ to $\boldsymbol{u}^{a}$ still holds under this inversion. Then, we will show that this simple, easy-to-implement method is yet very effective to achieve the good strategy.

*c.1)* $\Delta V_{PV,i}$: $\Delta V_{PV,i}$ determines the new operating point and thus the steady-state power output of the PV system. Since PV systems consist of identical PV panels connected in parallel with the same voltage $V_{PV,i}$ (as shown in Fig. 2), the higher rated power is, the more PV panels are connected, and thus, the higher $\Delta P_{PV,i}$ will be excited from the same $\Delta V_{PV,i}$.

Apart from rated power, solar irradiance $S$ is the other main variable that affects the power support capability of a PV system. The LUT surface plotted in Fig. 8 describes the relationship among $\Delta P_{panel}$, $\Delta V_{PV}$, and $S$. We can easily observe that, for the same value of $\Delta V_{PV}$, the $\Delta P_{panel}$ increases as the $S$ becomes higher. In other words, with the same $\Delta V_{PV,i}$, PV systems under higher $S$ will have more power change in each of their PV panel, and so in the total output.

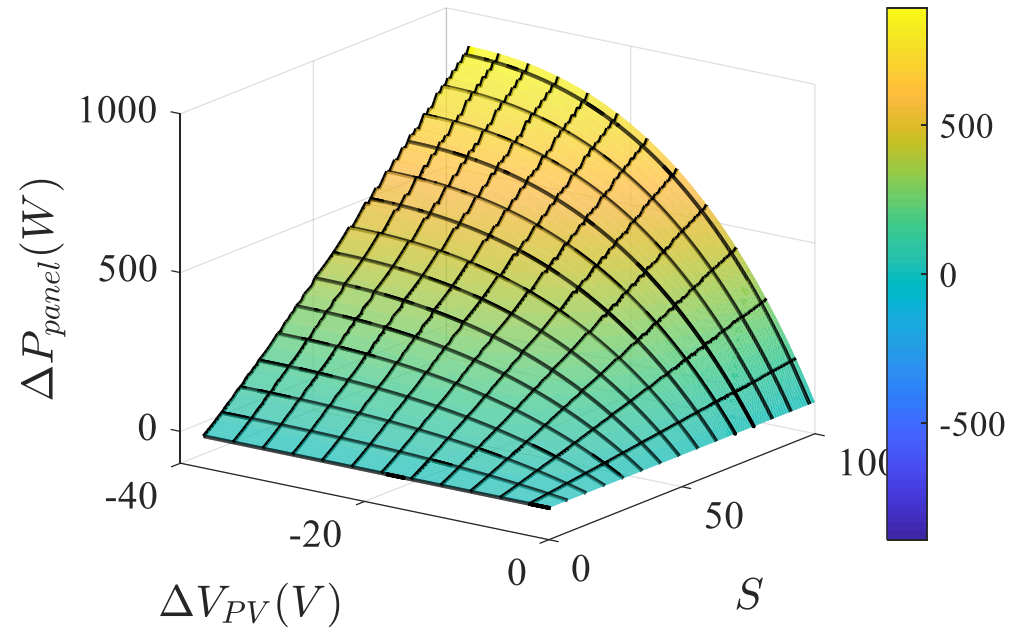

Fig. 8. The LUT surface of $g(\Delta V_{PV}, S, t)$ for a 5695 W PV panel at 26.85 °C

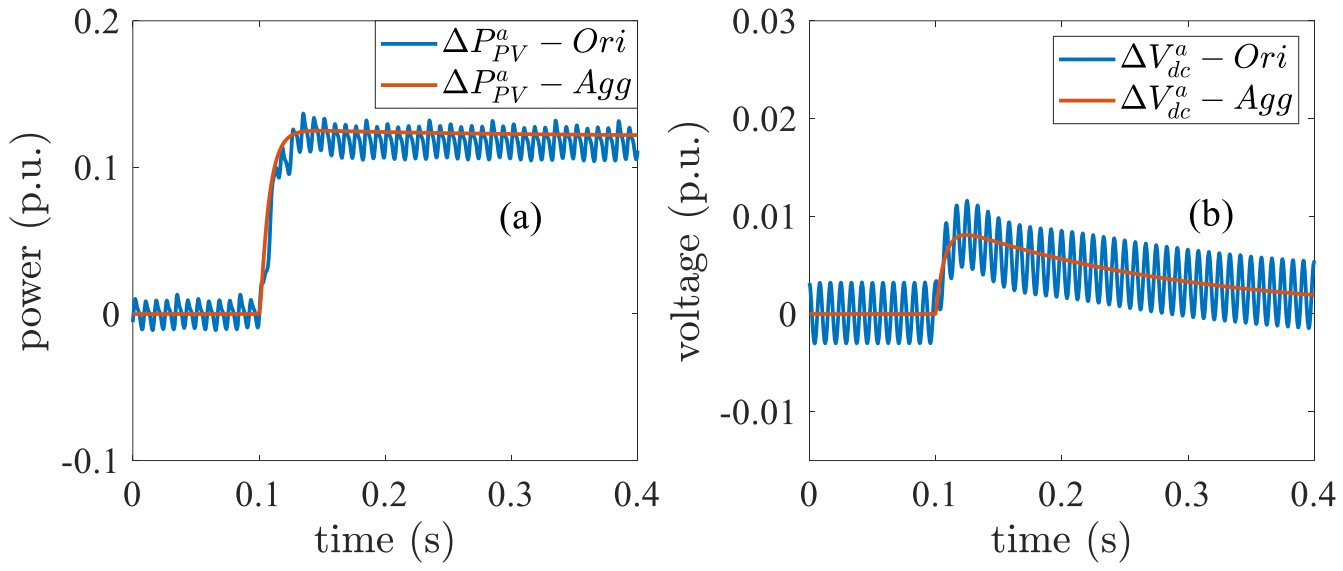

Fig. 9. When $\Delta V_{PV}^{a} = -12.775\ V$, (a) comparison of total increment power $\Delta P_{PV}^{a}$; (b) comparison of aggregate DC-link voltage $\Delta V_{dc}^{a}$

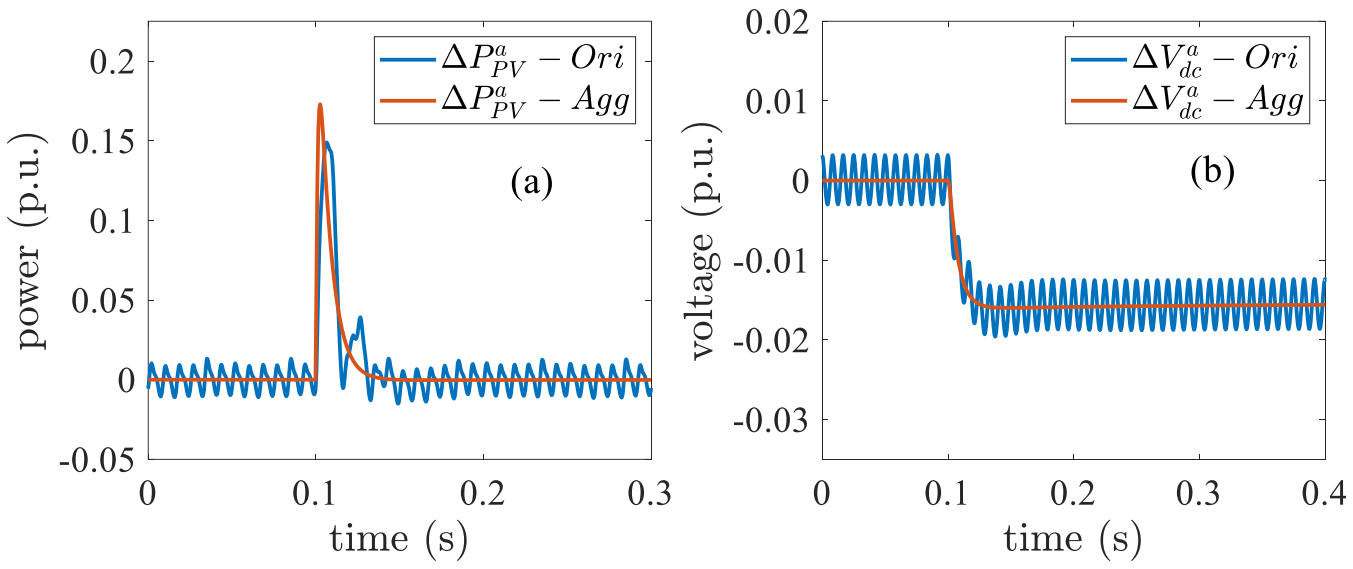

Fig. 10. When $\Delta V_{dc}^{ref,a} = -7.730\ V$, (a) comparison of total increment power $\Delta P_{PV}^{a}$; (b) comparison of aggregate DC-link voltage $\Delta V_{dc}^{a}$

*c.2)* $\Delta V_{dc,i}^{ref}$: The other control input $\Delta V_{dc,i}^{ref}$ adjusts the DC-link voltage to utilize power from the DC-link capacitor. According to (2), this power is proportionally related to the capacitance. We also know that the capacitance should be chosen in proportion to the PV system rated power, as described in (10). Therefore, with the same $\Delta V_{dc,i}^{ref}$, the PV system with higher rated power, and thus capacitance, will have more power released, or absorbed, from its DC-link capacitor.

## IV. Test Results

In this section, we first validate that the reduced-order aggregate model can effectively represent the overall dynamics of a group of distributed PV. Then, we demonstrate that the UIO-based tracking LQR, which is designed based on the aggregate model, is effective to enable frequency support from distributed PV during a disturbance event. Moreover, we also illustrate the effectiveness of proposed inversion method.

### A. Reduced-order Aggregate Model Validation

To validate the proposed reduced-order aggregate model, we first built an IEEE 34 node test feeder in MATLAB/Simulink to represent an area of distribution system. Then, 10 single-phase PV systems are connected to different locations. Each PV system is modeled as in Fig. 2. We use this test feeder as the benchmark system and compare its overall dynamics with that from the aggregate model. To represent the diversity of distributed PV systems, we choose different values for the solar irradiance, capacity, and control parameters, as listed in TABLE II. For the other basic parameters, we assume them to be the same for every PV system and use the values from [20].

TABLE II
Distributed PV Parameters with Variations

| | |
|---|---|
| $S_i$ | [100, 70, 65, 80, 85, 90, 92, 75, 83, 90] |
| $P_{r,i}$ (KW) | [200, 250, 220, 175, 200, 200, 250, 220, 175, 190] |
| $k_{P,i}$ | [10 15 20 20 17 10 15 20 20 17] |
| $k_{I,i}$ | [50 100 150 80 50 50 100 150 80 50] |

TABLE III
Individual and Aggregate Control Inputs for Validation Test

| | | | | |
|---|---|---|---|---|
| **Individual Input** | $\Delta V_{PV,i}$ | [-10, -12, -15, -11, -8, -5, -18, -20, -17, -10] | | |
| | $\Delta V_{dc,i}^{ref}$ | [-10, -5, -8, -7, -5, -4, -9, -11, -6, -12] | | |
| **Aggregate Input** | $\Delta V_{PV}^{a}$ | -12.775 V | $\Delta V_{dc}^{ref,a}$ | -7.730 V |

Since the aggregate model is primarily derived for designing frequency support control functions, we are particularly interested in the two relevant quantities: total PV system output power ($\Delta P_{PV}^{a}$) and aggregate DC-link voltage ($\Delta V_{dc}^{ref,a}$), following the changes in the control inputs. To test, the distributed PV are set to be operating at 85% of their MPPs (nominal de-loading condition). In the first test case, at $t =$ 1.5 s, we decrease the PV array voltages by different values as listed in TABLE III. In this case, we expect that the distributed PV to raise output power, as all the $\Delta V_{PV,i}$ are negative. For comparison, we calculate the corresponding aggregate control input ($\Delta V_{PV}^{a}$) and apply it to the aggregate model. We then compare the response of the aggregate model with the calculated aggregate response from the distributed PV in the benchmark system. The comparison results are provided in Fig. 9. In Fig. 9 (a), the total increment power from the aggregate model provides very good approximation to that obtained from the distributed PV. There is slight difference in the steady-state value which is caused by the LUT approximation we made in b.1). Fig. 9 (b) verifies that the aggregate DC-link voltage profile obtained from the aggregate model is very close to that of the distributed PV in benchmark system.

In the second test case, we perturb the other control input $\Delta V_{dc,i}^{ref}$. To extract power from DC-link capacitors, we decrease their voltage references by different values as listed in TABLE III. In Fig. 10 (a), fast aggregate power release is observed, and shortly after that, the power decreases to zero as the aggregate DC-link voltage shown in Fig. 10 (b) settles to its steady-state. Fig. 10 (a) and (b) show that the voltage and power profiles from the aggregate model are again very close to those from the distributed PV in benchmark system.

### B. Control Performance

To demonstrate the effectiveness of proposed control scheme, we consider the standard three-machine-nine-bus WECC system but modify its capacity down to 116 MVA ($S_b$) to represent a small-scale power system. Then we add four of the IEEE 34 node test feeders at bus 6, as shown in Fig. 11, with each feeder having 10 distributed PV. So, in this test system, we have 40 distributed PV, and their total capacity is 8.189

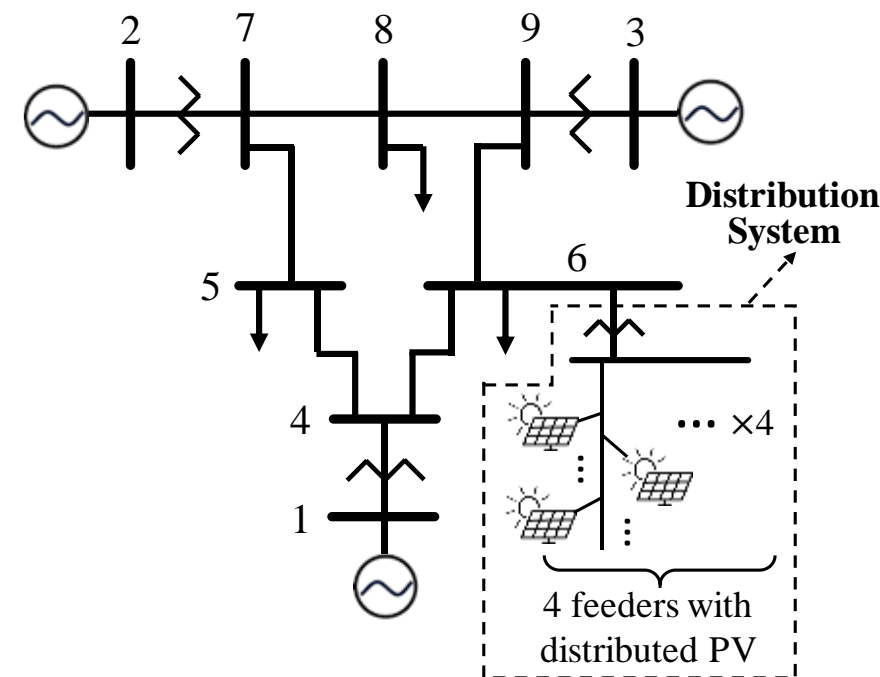

Fig. 11. Modified WECC 9-bus system with distributed PV

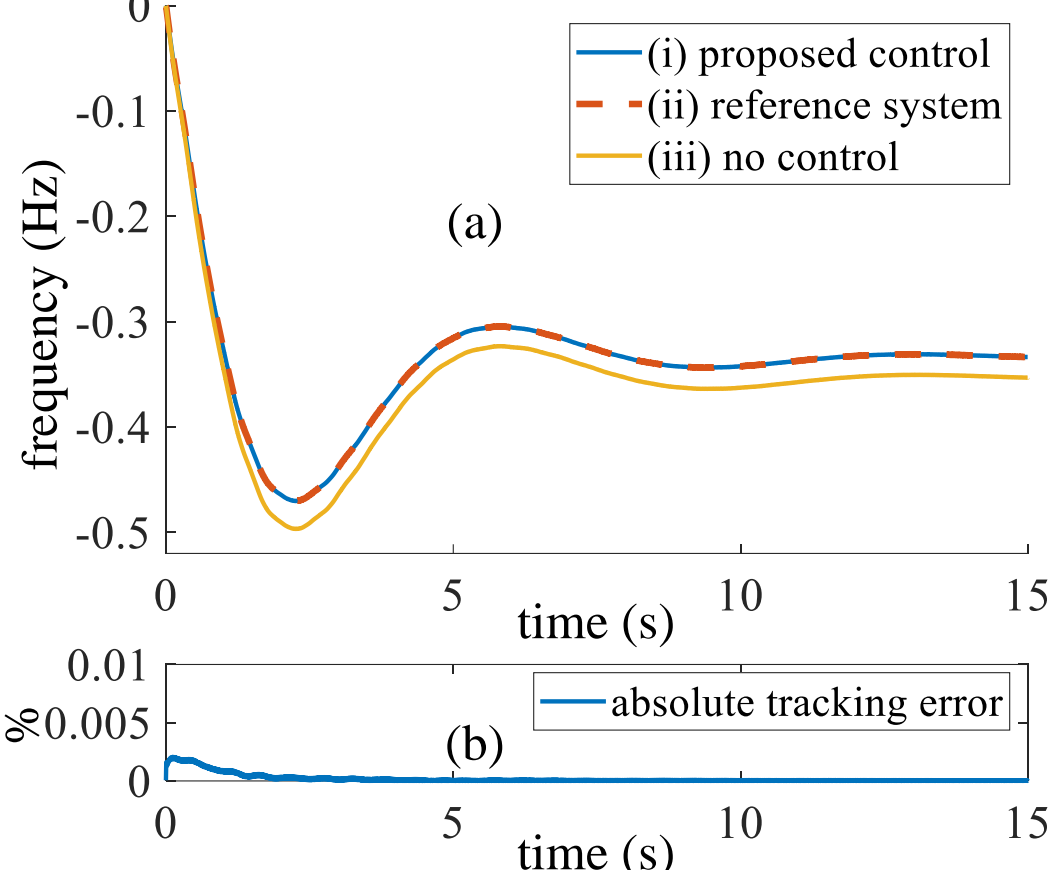

Fig. 12. (a) Frequency response comparison under load disturbance; (b) Absolute error between frequencies of (i) and (ii).

MW. The test system is built in MATLAB/Simulink where we conduct time-domain simulations with 5 μs time resolution for the following study.

To design the UIO-based tracking LQR, we follow the design steps in [20]. The same design parameters are chosen, except for $H^{ref}$ and $R^{ref}$ of the reference system as they should be adjusted based on the PV penetration level. We assume the distributed PV are added in replace of certain amount of synchronous generation. So, we choose $H^{ref}$ and $R^{ref}$ in the hope of restoring the frequency response capability to the original before adding PV. In this case, the desired inertia and droop constants for reference system are provided in TABLE IV, along with those for the original test system without proposed control.

For the test case, we apply a large load disturbance of 0.086 p.u. at bus 8. Fig. 12 (a) compares the frequency responses obtained from (i) system with proposed control scheme; (ii) reference system; and (iii) system with no frequency support control from distributed PV. From the figure we can clearly see that the frequency with proposed control effectively tracks that of the reference system frequency. In Fig. 12 (b), the absolute tracking error between frequencies from (i) and (ii) is given. It is shown that the tracking error is very small with the maximum value still being less than 0.005%. Fig. 12 (a) also indicates that the system frequency response is improved as desired, specifically in terms of nadir, RoCoF, and settling frequency, compared to that of the original system without control. Although the improvement is not very obvious, it is yet practically reasonable because the total capacity of distributed

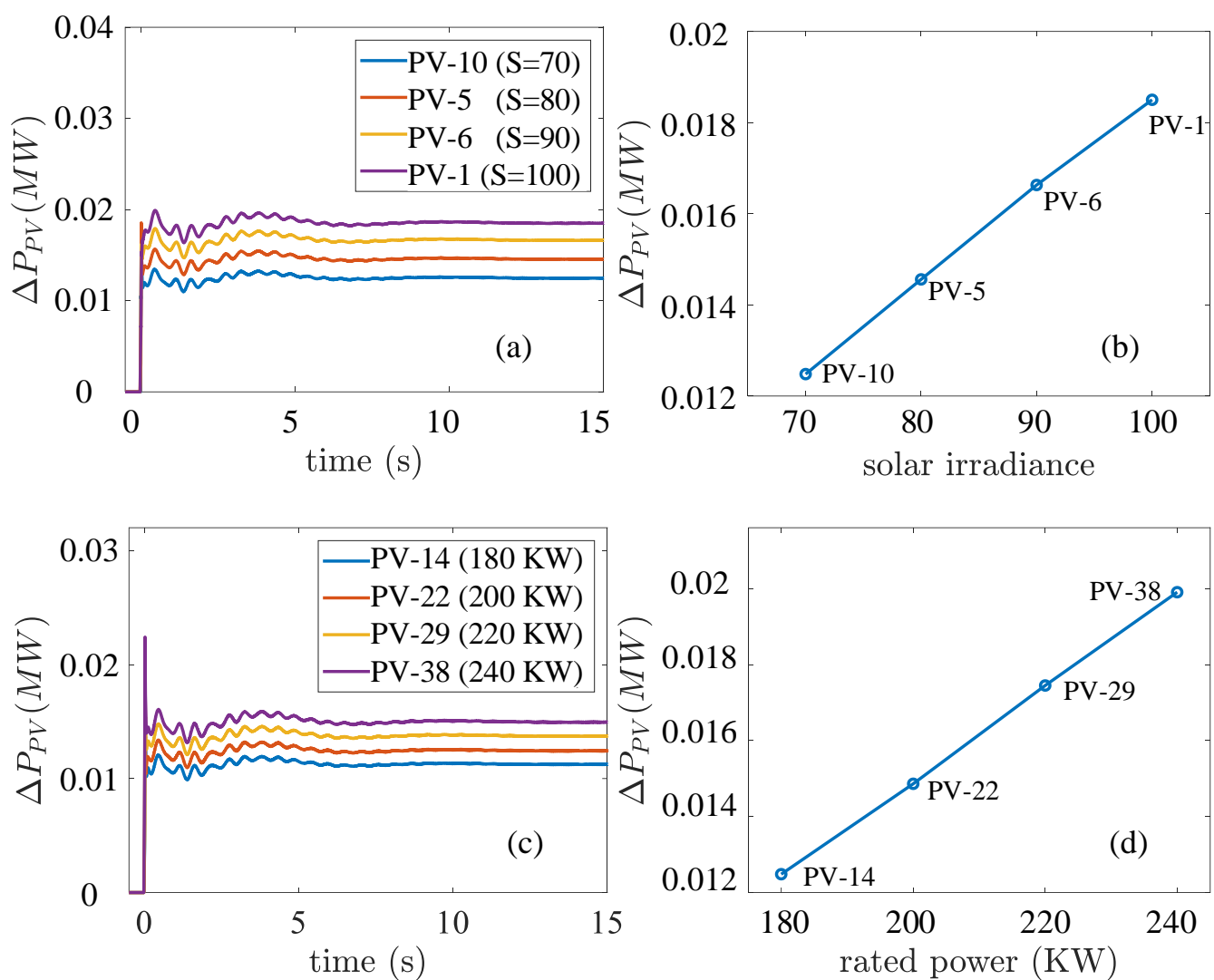

Fig. 13. For PV systems with the same rated power (200 KW), but different solar irradiance: (a) comparison of power responses; (b) correlation between steady-state $\Delta P_{PV}$ and $S$. For PV systems with the same solar irradiance (S=70), but different rated power: (c) comparison of power responses; (d) correlation between steady-state $\Delta P_{PV}$ and rated power.

TABLE IV
INERTIA AND DROOP USED IN CONTROL DESIGN

| | Test System | | Reference System | |
|---|---|---|---|---|
| **Inertia** | $H_g$ | 5.9746 s | $H^{ref}$ | 6.2365 s |
| **Droop** | $R_g$ | 8% | $R^{ref}$ | 7.66% |

PV is not expected to be very large in a power system. As a result, the choice of $H^{ref}$ and $R^{ref}$ should be conservative to match with PV's capacity and, correspondingly, frequency support capability.

Next, we want to verify the effectiveness of the proposed inversion method. To do this, we first select 4 distributed PV systems from the above simulation with the same rated power but different solar irradiance and plot their power responses in Fig. 13 (a). Per the control signal inversion in (34), they all share the same control inputs. As shown in the plot, the PV system with higher solar irradiance generates more power as expected. In the second case, we choose another set of 4 PV systems with the same irradiance but different rated power. Again, we can see from Fig. 13 (c) that the increment power outputs are positively correlated to the PV system capacities. In addition, as indicated in Fig. 13 (b) and (d), both of the positive correlations between steady-state $\Delta P_{PV}$ and solar irradiance $S$, $\Delta P_{PV}$ and rated power, are almost linear.

## V. CONCLUSION

Focusing on frequency support function of distributed PV, this paper proposes a novel supervisory control scheme, in which a frequency-tracking controller is designed to control a group of distributed PV as a whole. To this end, a reduced-order aggregate model is first developed to represent the overall dynamic behaviors of the distributed PV with diverse parameters. The derived model preserves the same order as that of one PV system model, and hence reduces the control design complexity from $\mathcal{O}(N)$ to $\mathcal{O}(1)$. Validation test indicates that the aggregate model provides a very close approximation of the real distributed PV dynamics in aggregate.

Based on the aggregate model, a frequency-tracking controller is improved for providing frequency support from distributed PV. An inversion method is also proposed as the critical supplement for implementing the controller in practice. With the inversion, individual control input for each distributed PV can be obtained from the aggregate input. Simulation results confirm the control performance of both frequency tracking and system frequency response improvement. It is also tested that the proposed inversion method provides an effective way of allocating the total control effort based on each PV's frequency support capability.